\documentclass[10pt,journal,letterpaper]{IEEEtran}

\usepackage[T1]{fontenc}
\usepackage{ae,aecompl}
\usepackage{epsfig}
\usepackage{amssymb}
\usepackage{amsfonts}
\usepackage{url}
\usepackage[dvips]{color}
\usepackage[usenames,dvipsnames]{xcolor}
\usepackage{amsmath}   

\newcommand{\ie}{{\it i.e.}}
\newcommand{\eg}{{\it e.g.}}

\newcommand{\argmax}[1]{\mathop{\mbox{argmax}}_{#1}}
\newcommand{\pdfarray}[2]{{ \left\{\begin{array}{ll} {#1}, & {#2} \\ 0, & o.w. \end{array}\right. }}
\newcommand{\Order}[1]{\mathcal{O}\left( {#1} \right)}

\definecolor{darkgreen}{rgb}{0,0.5,0}

\begin{document}

%

\title{Breathfinding: A Wireless Network that Monitors and Locates Breathing in a Home}

\author{Neal Patwari, Lara Brewer, Quinn Tate, Ossi Kaltiokallio, and Maurizio Bocca
\thanks{ N.~Patwari and M.~Bocca are with the Department
of Electrical and Computer Engineering, University of Utah, Salt Lake City, USA.  L.~Brewer is with the Department of Anesthesiology, University of Utah Health Sciences Center, Salt Lake City, USA.  Q.~Tate is with the School of Medicine, University of Utah, Salt Lake City, USA.  Ossi Kaltiokallio is with the Department of Automation and Systems Technology, Aalto University School of Electrical Engineering, 
Helsinki, Finland.  This material is based upon work supported by the National Science Foundation under Grant Nos. \#0748206 and \#1035565.  Contact email: npatwari@ece.utah.edu.}
}

\maketitle
\begin{abstract}
This paper explores using RSS measurements on many links in a wireless network to estimate the breathing rate of a person, and the location where the breathing is occurring, in a home, while the person is sitting, laying down, standing, or sleeping.  The main challenge in breathing rate estimation is that \emph{motion interference}, \ie, movements other than a person's breathing, generally cause larger changes in RSS than inhalation and exhalation.  We develop a method to estimate breathing rate despite motion interference, and demonstrate its performance during multiple short (3-7 minute) tests and during a longer 66 minute test.  Further, for the same experiments, we show the location of the breathing person can be estimated, to within about 2 m average error in a 56 m$^2$ apartment. Being able to locate a breathing person who is not otherwise moving, without calibration, is important for applications in search and rescue, health care, and security.
\end{abstract}



\section{Introduction} \label{S:Introduction}
This paper explores using standard wireless devices which measure only received signal strength (RSS) to monitor and localize the breathing of a person, who does not wear any device, in a residential environment.  In comparison, existing breathing monitoring methods either require sensors in contact with the body, or require specialized and expensive radar devices \cite{rivera2006multi,park2006single}.  The core of the idea of non-contact breathing monitoring via RSS is that some links' RSS values, on a particular channel, are highly sensitive to motion near the link line (the line between transmitter and receiver), and that those links can be used, in the absence of other motion, to estimate the person's breathing rate, and in addition, to estimate that a breathing person is near that link line.  From many links' RSS data, and from measurements on multiple frequency channels, we can estimate breathing rate and perform breathing localization. One of the core challenges is that these same links also 
experience rapid change due to other movement, for example, if a person's limb moves.  Our methods are robust to these normal occasional movements because it identifies when a sudden movement occurs, subtracts its effects out, and can still then estimate breathing rate in periods containing that motion.  Our work is the first, to our knowledge, to use RSS measurements to locate where that breathing is occurring.

A significant body of related research has shown that people moving people can be located in a building using a static wireless network that measures links' RSS values \cite{woyach06,zhang2007rf,patwari08b,zhang2009dynamic,kanso09b,wilson09a,chen11sequential}.  This can be done through walls \cite{wilson10see,zhao11noise,zheng2012through}, using a variety of statistics of the measured RSS.  However, state-of-the-art RSS-based non-cooperative localization methods require either: 1) \emph{motion}, that is, a person to be moving at least once during the course of the time in which measurements are collected \cite{wilson10see, youssef07,kaltiokallio2011}, or 2) \emph{empty-area calibration}, measurements of the RSS from the period of time when the area was cleared of any people \cite{patwari08b,wilson09a}.  Even systems sophisticated enough to adaptively learn to distinguish the statistics of RSS during crossing vs.~no crossing require some periods of both to occur \cite{zheng2012through}.  In an search and 
rescue operation or a building fire, people may be unconscious and thus motionless by the time emergency responders arrive, and thus a non-cooperative user localization system may not be able to locate them.  Although some work on robotic RF sensor networks has demonstrated the ability to image concrete structures \cite{mostofi2011compressive}, work has not yet shown that a motionless person could be located without prior empty calibration.  In this paper, we show a method that would locate a stationary but breathing person without an empty-area calibration, although we note that we do not test it through walls or in rubble.

As another application, using standard wireless networks for contact-free breathing monitoring may be useful for health care and wellbeing.  One may be able to use a wireless network deployed in a bedroom as a baby breathing monitor, or as part of an in-home sleep apnea diagnostic system.  
Elder care systems which track and monitor an elderly person's location and activities for abnormal changes could use breathing rates as additional context information.  As of yet, we have not developed algorithms to identify or distinguish the breathing rates of two people in the same room.  However, in these applications, breathing monitoring is most critical when the person is alone, and thus no other person is present to notice their apnea or their stoppage of breathing.  

Finally, in general, breathing rate provides a measure of a person's context, and context-aware computing systems could benefit.  Perhaps ``smart'' homes could also be ``empathetic'' and respond appropriately when your at-rest breathing rate is abnormally high.  A scary movie could adapt to the viewer's measured level of stress.




In the area of contact-free breathing monitoring, most systems use specialized radars.  Doppler radar devices can be used \cite{park2006single}, but slow and small breathing movements mean that the Doppler shift is very low unless the center frequency is very high.  As a result, commercial Doppler radar breathing monitoring products are expensive and out of reach for many applications. For example, the Kai Medical ``Continuous'' system \cite{kaimedical}, although not yet FDA approved, is said to be priced at \$2000.  Ultra wideband impulse radars can estimate breathing as well as heart rate \cite{rivera2006multi}.  Radar propagation losses are of order $d^4$ in free space \cite{rappaport2001com}, as opposed to $d^2$ for transmission \cite{stutzman}, and thus radar devices either have limited range or high transmit power.  Regarding the latter, UWB devices are severely limited in transmit power by regulation, and thus compliant devices have a low range.  We believe that use of a standard wireless network 
offers a low-cost alternative for residential breathing monitoring.  

Our initial work was presented in \cite{patwari11breathing}, which introduced RSS-based breathing monitoring,  presenting algorithms for estimating breathing rate and detecting the absence of breathing. The methods were experimentally verified on a subject in a hospital bed, using transceivers placed within centimeters of the subject's body.  The report shows that links in a deep fade are the ones most likely to measure changes when the person breathes.  Our breathing rate estimator is compared to the method of \cite{patwari11breathing} in Section \ref{S:Monitoring}. 

This paper makes several advancements:
\begin{enumerate}
 \item This paper is the first to demonstrate that breathing can be monitored using wireless networking devices meters, as opposed to centimeters, away from the person.  In addition, we use measurements on multiple frequency channels, and quantify the improvement possible compared to single-channel RSS measurements in Section \ref{S:NumberOfChannels}.  Our testbed and experimental setups are described in Section \ref{S:Experiments}.  
 \item This paper presents an algorithm to estimate and avoid motion interference, and as a result, estimate breathing rate even during windows of time in which there is motion in the environment.  The subject in \cite{patwari11breathing} was completely stationary, except for an expanding and contracting chest.  When the person moves, for example, coughs, scratches an itch, moves an arm or leg, or changes position in bed, the method of \cite{patwari11breathing} becomes unable to estimate breathing rate for a period of time (\eg, 30 s). 
Fundamentally, this is because the links most sensitive to chest motion are also those most sensitive to other motion.  Our method shows dramatic breathing rate estimation performance improvements.  Using 30 seconds of data, our method's breathing rate estimates have average error of 1.0 bpm, compared to 1.7 bpm for \cite{patwari11breathing}.  Our breathing rate estimation algorithm and experimental results are presented in Section 
\ref{S:Monitoring}.
 \item Third, the paper is the first to develop and test an algorithm to locate a breathing but otherwise stationary person.  Without any empty-room calibration or training period, we estimate an image map of breathing intensity in the deployment area, and locate the breathing from the highest point in the image.  We see average localization errors of about 2 m. Our localization algorithm and experimental results are presented in Section \ref{S:Localization}.  
\end{enumerate}
Finally, we conclude and discuss future research directions in Section \ref{S:Conclusion}.





\section{Experiments} \label{S:Experiments}

In this section, we describe the testbed network, the residential environment in which the tests are conducted, and the testing procedure. 

\subsection{Network}


In general, we suppose there is a network of $S$ transceiver nodes which can operate on one of $C$ frequency channels.  Link $l$ has transmitter (TX) $T_l$, receiver (RX) $R_l$, and frequency channel $F_l$, \ie, each TX / RX / channel combination is a different logical link. In our experiments, $L=C S(S-1)$ because all links are connected, but full connectivity is not necessary for breathing monitoring.  

We assume that each link's RSS is measured at regular intervals with period $T$ by sending and receiving a packet.  Note that newborns have the highest breathing rates, near 37 breaths per minute (bpm), or 0.62 Hz, and adults at rest breathe near 10-14 bpm, or 0.23 Hz \cite{murray86}.  To sample at the Nyquist rate, sufficient for a newborn monitoring application, our sampling rate would need to be more than 1.24 Hz.  We denote the $n$th RSS on link $l$ as $r_l[n]$, in dB units.  Future work could extend the presented algorithms for irregularly sampled data.

In our experiments, we use the TI CC2531 dongle node \cite{tidongle}, an IEEE 802.15.4 compliant radio.  The CC2531 transmits with transmit power 4.5 dBm in the 2.4 GHz ISM band, and can select from one of 16 frequency channels.


A multi-channel TDMA protocol is used in which each node transmits in sequence, and after all nodes' slots, nodes switch synchronously to the next frequency channel \cite{kaltiokallio2012follow}.  A node's packet transmission includes its node ID and the RSS of the most recent packet received from each other node.  A sink node that overhears all packets stores the data for post-processing.

\subsection{Apartment Experiment}

In our first experiment, the \emph{apartment experiment}, we deploy $S=33$ nodes, each at a height of 0.9 meters.  Our protocol uses $C=4$, specifically, 802.15.4 channel numbers $15$, $20$, $25$, and $26$, selected to overlap the least with WiFi.  Each link is measured each $T=428$ ms.  This is a sampling rate of 2.3 Hz. 

\begin{figure}[tbhp]
\centerline{ \epsfig{figure=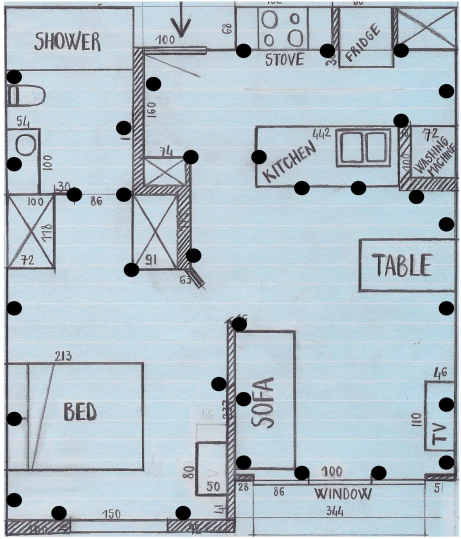,width=2.4in} }
\caption{Floor plan of 7 $\times$ 8 m apartment, deployed nodes ($\bullet$), and dimensions in cm.}
    \label{F:ApartmentMap}
\end{figure}

\begin{figure}[hbtp]
\centerline{ \epsfig{figure=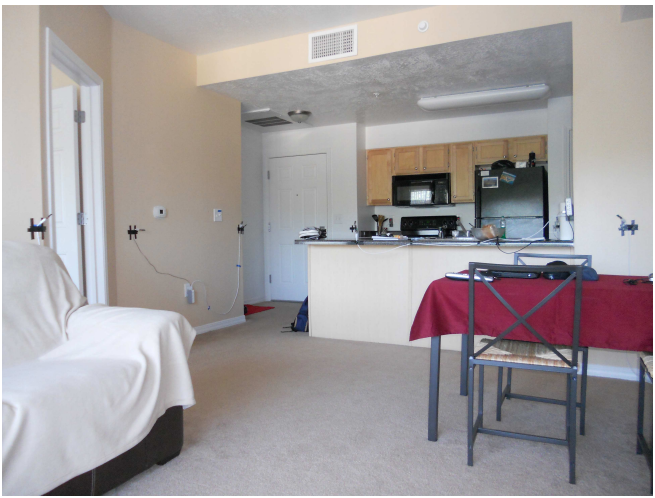,width=3.4in} }
\caption{Living and kitchen areas in apartment, with nodes seen taped to the wall at height 0.9 m.}
    \label{F:ApartmentPhoto}
\end{figure}

The testbed network is deployed in the apartment shown in Figure \ref{F:ApartmentMap}.  Node locations are indicated with circles on the map.  The apartment is on the third floor of a fully-occupied six story building in downtown Salt Lake City, Utah.  It is an actual person's residence, and is furnished, as can be seen from the photo of the living and kitchen areas in Figure \ref{F:ApartmentPhoto}.  Tests are conducted on a Saturday between noon and 1:00 pm.  During the tests, other people are sometimes heard in the hallway outside of the apartment and in neighboring apartments.

As a large apartment building in a metropolitan area, there are many sources of 2.4 GHz interference. The apartment has its own wireless router, and more than ten strong WiFi access points are observed, and although the channels are chosen to have the least overlap with WiFi, they are not perfectly orthogonal.  Microwave ovens and cordless phones also exist in the nearby environment. 

\noindent{\bf Tests:} Five tests are conducted.  In each, the experimenter stays in a single location for 3-7 minutes and maintains maintains a constant breathing rate of 10 bpm using a metronome.  The experimenter is instructed to sit or stand still at the location, but to move as required to be comfortable, \ie, move an arm or leg or change position.  The experimenter made some movement, on average, twice per minute.  In the tests, the experimenter is located: (1) standing in the kitchen; (2) sitting by the dining table; (3) sitting on the sofa; (4) laying in bed; and (5) sitting on the toilet.

\subsection{Nap Experiment} \label{S:Experiment2}

\begin{figure}[tbhp]
\centerline{ \epsfig{figure=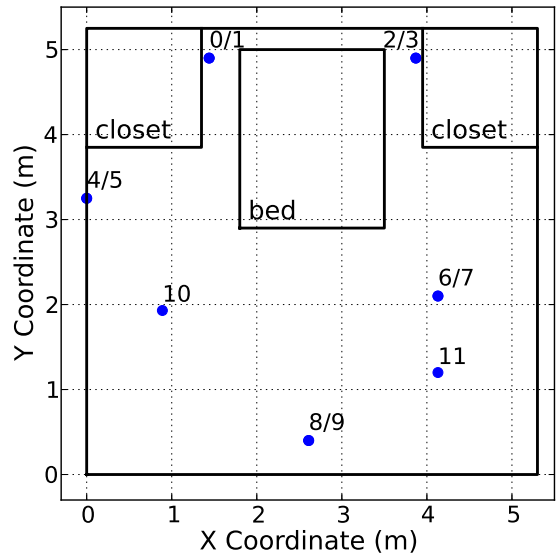,width=3.4in} }
\caption{Master bedroom floor plan, for ``nap'' experiment.  Where two nodes are at the same $(x,y)$ coordinate on this 2D map, the higher node number is listed first.}
    \label{F:map_nap_experiment}
\end{figure}

Our second experiment, the \emph{nap experiment}, is conducted in the master bedroom on the second floor of a house, as shown in Figure \ref{F:map_nap_experiment}.  $S=12$ nodes are deployed in a 5.3 m by 5.3 m area, at power outlet locations or in power strips in the room.  We placed two nodes per power outlet, one on the floor and the other at a height of about 0.48-1.65 m.  The height of the high nodes varied because they are placed on top of furniture (bedside table, dresser, table, etc.).  In this experiment, $C=5$ channels are measured, specifically, IEEE 802.15.4 channels 11, 15, 18, 22 and 26.  Here, the sampling period was 179.6 ms, for a sampling rate of 5.56 Hz.

There is also WiFi interference at this location, with one wireless router on the first floor below the bedroom.  A laptop at location (2m, 0.5m) records data and sends it over WiFi to a server, ensuring some level of WiFi interference to the breathing monitoring system.

A single 66-minute long test is conducted.  The experimenter starts collecting data on the laptop, walks to and climbs into the bed, and takes a nap.  In contrast to the previous experiments, no ground truth is known for the experimenter's breathing rate, as no separate breathing monitoring sensor is attached to the person, and the person cannot maintain a constant breathing rate while sleeping.  After time 65 minutes, the person wakes up, gets out of bed and stops the data recording.  The person moved at least once during sleep, since it was noted that he did not wake up in the same position that he went to sleep.

\section{Breathing Rate Estimation} \label{S:Monitoring}

\subsection{Basic Method}

In \cite{patwari11breathing}, an RSS-based breathing rate estimator is proposed which, in short, calculates the power spectral density (PSD) over each link using the most recent $N$ samples, sums the PSD over $L$ links, and estimates the breathing rate as the frequency at the maximum of the sum PSD.  Mathematically,
\begin{equation} \label{E:PSD}
\hat{f} = \argmax{f_{min} \le f \le f_{max}} \sum_{l=1}^L \left| \sum_{n=i-N+1}^{i} y_l[n] e^{-j 2\pi f T n}\right|^2,
\end{equation}
where $f_{min}$ and $f_{max}$ are the minimum and maximum breathing rates which the system is set to allow, $i$ is the current time, $j=\sqrt{-1}$, and $y_l[n]$ is the change in RSS compared to the mean, defined as
\begin{equation}\label{E:MeanRemoval}
y_l[n] = r_l[n] - \bar{r}_l,
\end{equation}
where $\bar{r}_l$ is the average RSS on link $l$.  We call $y_l[n]$ the \emph{RSS signal} because the mean RSS carries no information about breathing, and in fact, a strong DC component overwhelms any other frequency contained in the data.  In \cite{patwari11breathing}, DC removal is performed using a 7-tap IIR filter.  However, we find that setting $\bar{r}_l = \frac{1}{N} \sum_{n=i-N+1}^{i} r_l[n]$, \ie, the window average, has superior performance over a wide variety of experiments we have conducted.  When comparing our method, to be fair to the spirit of the method in \cite{patwari11breathing}, we use this window average in (\ref{E:MeanRemoval}).

\subsection{Breakpoint Method}

The challenge faced by the RSS-based breathing monitoring in general, and the basic method in particular, is that the links which best measure the person's chest movement are also those that best measure other motion in the environment.  This other motion, which we term ``motion interference'', can hide the breathing-induced changes.  
From our observation, motion interference and breathing cause notably different types of changes.  While breathing causes slow, periodic RSS changes, motion interference typically causes sudden changes.  For example, a person who rolls over in bed, or moves a foot, causes fast RSS changes during their movement.  When the person stops moving, a link's RSS typically settles to an average value different than before the movement.  If such a shift occurs during the $N$-sample window, subtracting the mean as in (\ref{E:MeanRemoval}) will not suffice to remove the strong low-frequency component in the RSS signal.

Instead, we propose to detect each time index during which a sudden RSS change occurs.  We call these detected time indices ``breakpoints''.  At each breakpoint, the system is allowed to forget the previous link RSS averages and calculate new averages.  By doing so, we are able to perform more accurate mean removal, and thus make breathing rate estimation more robust to motion interference.

The T-test is an appropriate statistical detection test to perform this task.  The T-test is a generalized likelihood ratio test (GLRT) of the change in mean between two groups of data \cite{kay1993detection}, in our case, one group of RSS signal samples prior to the time-under-test (the sample index at which we want to detect a change in mean), and another same-duration group of data after the time-under-test. Specifically, the group t-score is,
\begin{equation} \label{E:Tscore}
  \tau_l[n] = \frac{ \grave{r}_l[n] - \acute{r}_l[n]}{ \max\left\{ \epsilon, \sqrt{ (\grave{\sigma}^2_l[n] + \acute{\sigma}^2_l[n]) / Q} \right\}},
\end{equation}
for each link $l$ and time $t$, where $\grave{r}_l[n]$ and $\acute{r}_l[n]$ are the average of the $Q$ samples of $r_l[m]$ before and after time $n$, respectively, $\grave{\sigma}^2_l[n]$ and $\acute{\sigma}^2_l[n]$ are the sample variances of $r_l[m]$ before and after time $n$, respectively, and $\epsilon>0$ is used to prevent division-by-zero.  We use $\epsilon=0.5$ in our experiments.

Next, at each time $n$ we compute the root-mean-squared (RMS) average of $\tau_l[n]$ over all links $l$,
\begin{equation}
  \tau_{RMS}[n] = \left( \frac{1}{L} \sum_{l=1}^L \tau^2_l[n]  \right)^{\frac{1}{2}}.
\end{equation}
Time $n$ is a breakpoint index if $\tau_{RMS}[n] \ge \gamma$, where $\gamma$ is a predetermined threshold.  For any time window, we also consider the starting time index and ending time index as breakpoints.  At the current time $i$, with a window length $N$, the starting and ending time indices are $i-N+1$ and $i$.

Breakpoints are used to remove the mean from the raw RSS as follows:
\begin{equation}\label{E:MeanRemovalBreakpoint}
y_l[n] = r_l[n] - \tilde{r}_l[n],
\end{equation}
where $\tilde{r}_l[n]$ is the average of $r_l[m]$ for all $m$ such that $b_p \le m < b_f$, where $b_p$ is latest breakpoint before or at time index $n$, and $b_f$ is the earliest breakpoint after time $n$.  Essentially, the mean model $\tilde{r}_l[n]$ is piecewise constant, with transition times at each breakpoint. After performing mean removal in (\ref{E:MeanRemovalBreakpoint}), we use our new RSS signal $y_l[n]$ in (\ref{E:PSD}) to perform breathing rate estimation.

\begin{table}[tbp]
\begin{center}
\begin{tabular}{|cc|}
\hline
\bf Parameter & \bf Value \\
\hline
Window Duration, $N$ & 70 (30 sec) \\
Breathing Rate Min., $f_{min}$ & 0.1 Hz (6 bpm) \\
Breathing Rate Max., $f_{max}$ & 0.4 Hz (24 bpm) \\
Nodes, $S$ & 33 \\
Channels, $C$ & 4 \\
Sampling Period, $T$ & 0.428 sec \\
T-test Window, $Q$ & 14 (5 sec)\\
RMS T-score Threshold, $\gamma$ & 0.8 \\
\hline
\end{tabular}
\end{center}
\caption{Breathing Rate Estimator and Experiment Parameters. } \label{T:Parameters}
\end{table}

\subsection{Apartment Experiment Results}

\begin{figure}[hbtp]
\centerline{ \epsfig{figure=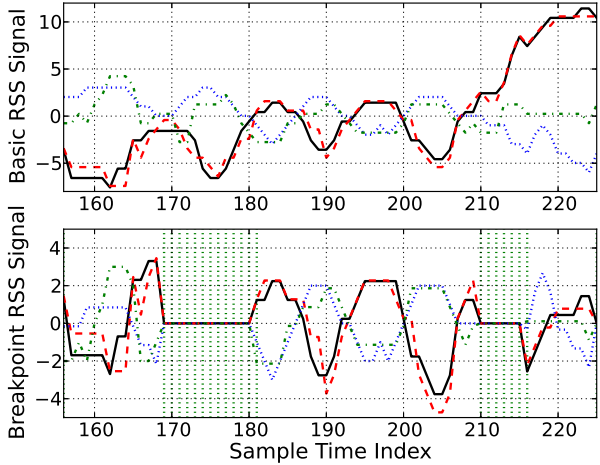,width=3.6in} }
\caption{The mean-removed RSS signal $y_l[n]$ for four links that best show breathing during `sofa' experiment, using (top) basic and (bottom) breakpoint methods, with estimated breakpoints (\textcolor{darkgreen}{\bf :}).}
    \label{F:RSSSignal_2}
\end{figure}

To compare the basic and breakpoint methods, we describe an example taken from a 30-second window during the `sofa' experiment.  Methods are tested using parameters given in Table \ref{T:Parameters}.  Figure \ref{F:RSSSignal_2} shows the mean-removed RSS signal $y_l[n]$ for the basic and breakpoint methods, for four links $l$ that are the four best in terms of signal amplitude at the true breathing frequency.   In the basic method, $y_l[n]$ drifts upward, particularly around $n=213$.  Many links (including those not shown) exhibit a shift in mean here, as well as near $n=175$.  The breakpoint method, using a threshold $\gamma=0.8$, calculates a high RMS t-score near these two times, and detects several time indices near 213 and 175 as breakpoints.  By removing the mean in between each pair of breakpoints, the signal $y_l[n]$ no longer has a drifting mean in the breakpoint method.  Note that some ranges are reduced to zero because breakpoints are dense, and removing the average from a single sample results in 
a zero value.

\begin{figure}[tbp]
\centerline{ \epsfig{figure=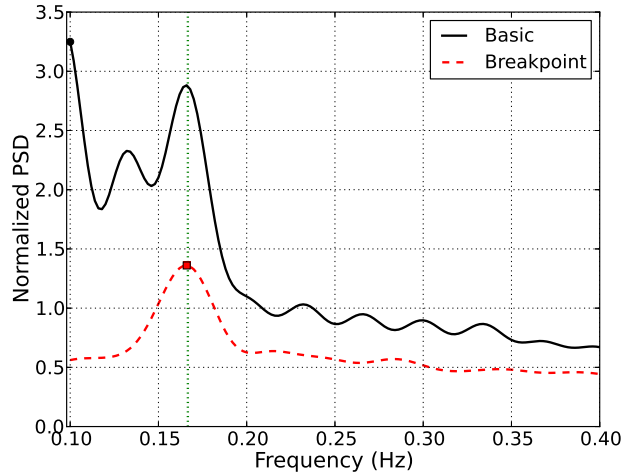,width=3.6in} }
\caption{Normalized average PSD from (\ref{E:PSD}) using basic and breakpoint methods and data from Figure \ref{F:RSSSignal_2}, estimated breathing rate ($\bullet$ and $\square$), and true breathing rate (\textcolor{darkgreen}{\bf :}).}
    \label{F:PSD_2}
\end{figure}

Figure \ref{F:PSD_2} shows that the breakpoint method enables correct breathing rate estimation when the basic method is unable.  The basic method's PSD is shown to be highest at the lowest breathing rate, 0.10 Hz, and thus $\hat{f} = 0.10$ Hz.  The PSD for the breakpoint method is reduced at all frequencies, but has a clear maximum at 0.166 Hz, very close to the true 0.167 Hz breathing rate.

Next we test all 30 second windows for all test locations, moving the window 5 seconds each time.  
We first report the fraction of estimates $\hat{f}$ which are \emph{acceptable}, which we arbitrarily define as being within 3.0 bpm of correct. Second, we report the average frequency error over all windows.  


\begin{table}[tbp]
\begin{center}
\begin{tabular}{|l|cc|cc|}
\hline
\bf          & \multicolumn{2}{c|}{\bf \% Acceptable} & \multicolumn{2}{c|}{\bf Avg.~Err.~(bpm)} \\
\bf Test Loc &  \bf Basic & \bf Breakpt & \bf Basic & \bf Breakpt \\
\hline
 Sofa       & 71\%   & 94\%  & 1.23 & 0.348 \\ 
 Table      & 81\%   & 89\%  & 1.14 & 0.831 \\ 
 Kitchen    & 58\%   & 91\%  & 1.81 & 0.859 \\
 Bathroom   & 71\%   & 97\%  & 1.29 & 0.339 \\
 Bed        & 25\%   & 33\%  & 2.95 & 2.63  \\
\hline
\bf Average & \bf 61\%   & \bf 81\%  & \bf 1.69 & \bf 1.00 \\
\hline
\end{tabular}
\end{center}
\caption{Basic vs.~breakpoint breathing rate estimation: percent of estimates ``acceptable'', and average error. } \label{T:FreqResults}
\end{table}

Results are shown in Table \ref{T:FreqResults}.  Over all tests, the basic method obtains an acceptable rate estimate 61\% of the time.  The breakpoint method achieves 81\% acceptable estimates, approximately cutting in half the number of `unacceptable' rate estimates.  With the breakpoint method, four of five tests have acceptable rates above 89\%, whereas the basic method had none.  The breakpoint method reduces the average breathing rate error to 1.0 bpm, a 41\% reduction from the basic method.  Primarily, the reduction in average error comes from eliminating large errors -- both methods typically have less than 0.3 bpm error when the rate estimate is ``acceptable''.  For reference, an end-tidal CO$_2$ meter, the gold standard breathing rate monitor in hospitals \cite{cook2011major}, is accurate to $\pm 1$ bpm.

The bed test has particularly poor performance.   We note that if we had used the breakpoint method with $\gamma = 0.5$ (instead of 0.8), the bed test would have had 73\% acceptable estimates (vs.~33\%), however, the table test would fare worse with the lower threshold.  Future work should address adaptive methods to decide upon breakpoints.

\subsection{Nap Experiment Results}
The data from the nap experiment is processed with the breakpoint method, using the same parameters as given in Table \ref{T:Parameters}.  

\begin{figure}[htbp]
\centerline{ \epsfig{figure=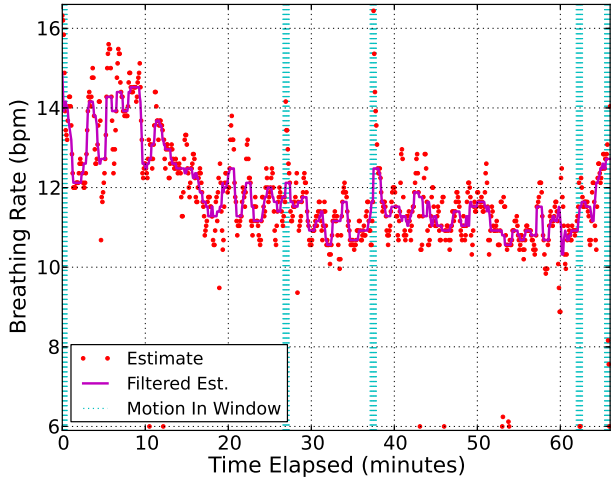,width=3.7in}}
\caption{Breakpoint method nap experiment breathing rate estimates based on 30 s data window (\textcolor{red}{$\cdot$}), 90 s median (\textcolor{Mulberry}{-----}), and breakpoints (\textcolor{green}{$\vdots$}) which indicate periods when movement is detected.  }
    \label{F:nap_fHat_breakpts_median_vs_time}
\end{figure}

As discussed in Section \ref{S:Experiment2}, no ground truth is available, however, there are features of the breathing rate estimates which match our understanding of the human body during sleep.  It has been shown that breath rate decreases during the transition from the awake state to stage two of sleep \cite{white1985metabolic}. This is primarily because there is a downward trend in energy expenditure during the onset of sleep \cite{white1985metabolic,katayose2009metabolic}.  The breath rate in Experiment 2 decreases approximately 20\% in the first 20 minutes, which correlates with published data showing a measured metabolic rate decrease of approximately 20\% during the first 20 minutes of the onset of sleep \cite{katayose2009metabolic}. During sleep stages 2, 3/4 and REM (rapid eye movement), the ventilation and metabolic rate are known to become relatively stable \cite{white1985metabolic,douglas1982respiration}, which is consistent with what we observed in our nap experiment estimates. However, moving 
during sleep increases energy expenditure and thus could temporarily increase breathing rate \cite{white1985metabolic}, as we see clearly at time 38 minutes, just after a period in which motion is detected by the breakpoint method. At the end of the nap experiment, the breath rate increased as the person awoke without an alarm. This is consistent with published data showing increased movement and metabolic rate during the transition to wakefulness \cite{katayose2009metabolic}.

In a period containing motion that is not detected and removed by the breathing rate estimation algorithm, the rate estimate tends to rail to $f_{min}$.  This is because a single step function, when the step is large compared to other signals, would be estimated to be evidence of one cycle in the measurement period.  For a 30 second window, this would be a rate of 2 bpm.  Since $f_{min} = 6$ bpm in these experiments, the estimator will return a value of 6 bpm in this case.  Step functions are not removed when using the basic algorithm, and in some instances, the breakpoint method may not detect a step function if the amplitude of the change is small compared to the variances before and after the jump.  Although we are not certain of the sleeping person's true breathing rate, we quantify the number of rate estimates at or very near 6 bpm as a measure of the ability of the method to remove normal (non-breathing) motion from the breathing-induced signal.  In this experiment, using the breakpoint method, there 
are fourteen breathing rate estimates below 9 bpm, five of which are during identified periods of motion.  Thus out of 793 window $\hat{f}$ estimates, only nine (1.1\%) are very low and are during periods when no motion is detected.  

\begin{figure}[htbp]
\centerline{ \epsfig{figure=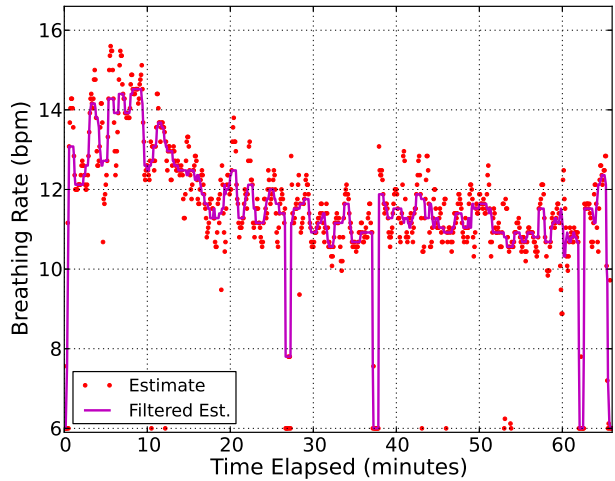,width=3.7in}}
\caption{Basic method nap experiment breathing rate estimates based on 30 s data window (\textcolor{red}{$\cdot$}), and 90 s median (\textcolor{Mulberry}{-----}).  }
    \label{F:nap_fHat_basic_median_vs_time}
\end{figure}

In comparison, we show results for the basic mean-removal method in Figure \ref{F:nap_fHat_basic_median_vs_time}.  When using the basic method, two metrics are worse.  First, there are 48 estimates with rate below 9 bpm.  Since there are several of these low estimates in a row, the median filter also falls below 9 bpm during periods that we believe (from the breakpoint method results) contain movement.  In contrast, the median filter in the breakpoint method never falls below 9 bpm.  Not only are the basic method rate estimates presumably incorrect, in addition, there is no ``motion'' label provided to indicate to the application that there was likely motion in that period.  As a result, there is no way to automatically discard the poor rate estimates returned by the basic method.

\section{Breathing Localization} \label{S:Localization}

In this section, we describe a method to compute a ``map'' of where the breathing, at the estimated rate, is occurring.  From the map we estimate the coordinate of the breathing experimenter within the apartment.

Variance-based radio tomographic imaging (VRTI) uses the variance of RSS on links in a wireless network to identify the location of a moving person in a building \cite{wilson10see}.  However, a breathing but stationary person does not show up in a VRTI image, as the image is more affected by noise than by the small changes caused by breathing.  In this section, we adapt the approach from \cite{wilson10see} for breathing mapping and localization, and we show using the experimental data that an approximate location of the breathing person can be determined.

\subsection{Method}

We use the PSD at the estimated breathing rate for each link as an input to our method.  For link $l$, at time $i$, the PSD of link $l$ at $\hat{f}$ is,
\begin{equation}
  v_l = \left| \sum_{n=i-N+1}^{i} y_l[n] e^{-j 2\pi \hat{f} T n}\right|^2.
\end{equation}
The vector of all link values is $v = [v_1, \ldots, v_L]^T$.

We want to estimate an image of breathing amplitude vs.~space, denoted vector $x$, where $x_k$ represents the quantity of breathing energy coming from pixel $k$.  As in \cite{wilson10see}, we assume that $v$ is a linear combination of $x$ via a weighting matrix $W$ plus noise $\eta$,
\begin{equation} \label{E:linearCombo}
   v = Wx + \eta
\end{equation}
where $W$ is defined as having $(l,k)$ element given by,
\begin{equation} \label{E:WeightModel}
   W_{l,k} = \pdfarray{\frac{1}{P_{l}}}{\frac{\| z_{T_l} - p_k \| + \| z_{R_l} - p_k \|}{\| z_{T_l} - z_{R_l} \| + \lambda_e} \le 1},
\end{equation}
where $z_{T_l}$ and $z_{R_l}$ are the coordinates of TX and RX for link $l$, respectively, $p_k$ is the coordinate of pixel $k$, $\lambda_e$ is the ellipse size parameter, and $P_{l}$ is the number of non-zero weights for link $l$.  Essentially, $P_{l}$ normalizes the weight so that the total weight of each link is 1.  

\begin{table}[tbp]
\centering
\begin{tabular}{|lc|}
\hline
\bf Parameter & \bf Value \\
\hline
Pixel Width, $\delta_p$ & 0.2 m \\
Pixel Variance, $\sigma_x^2$ & 2 \\
Correlation Distance, $\delta$ & 2 m \\
Ellipse Size Parameter, $\lambda_e$ & 1 m \\
\hline
\end{tabular}
\caption{Breathing mapping and localization parameters } \label{T:LocParameters}
\end{table}

Solving for $x$ in (\ref{E:linearCombo}) given measurement $v$ is ill-posed.  We assume the image vector has covariance matrix $G$, with $(k,m)$ element $G_{k,m} = \sigma_x^2 e^{-\| z_k - z_m \|/\delta}$, where $\sigma_x^2$ is the variance of any element of $x$, and $\delta$ is the correlation distance.  The regularized least squares solution for $x$ is thus,
\begin{equation} \label{E:inversion}
\hat{x} = \Pi v, \quad \mbox{where } \Pi = \left(W^T W + G^{-1}\right)^{-1} W^T.
\end{equation}
Note $\Pi$ must be computed only once.  The real-time computation of the image requires only one matrix multiply, of $\Order{LP}$ multiplies and adds, where $P$ is the number of pixels.

From the image, the coordinate of the pixel with maximum value in $\hat{x}$ is used as the location estimate for the (single) breathing person.  Future work must address localization in the multi-person case.

\subsection{Apartment Experiment Results}

\begin{figure*}[bth]
\centerline{ 
             (a) \epsfig{figure=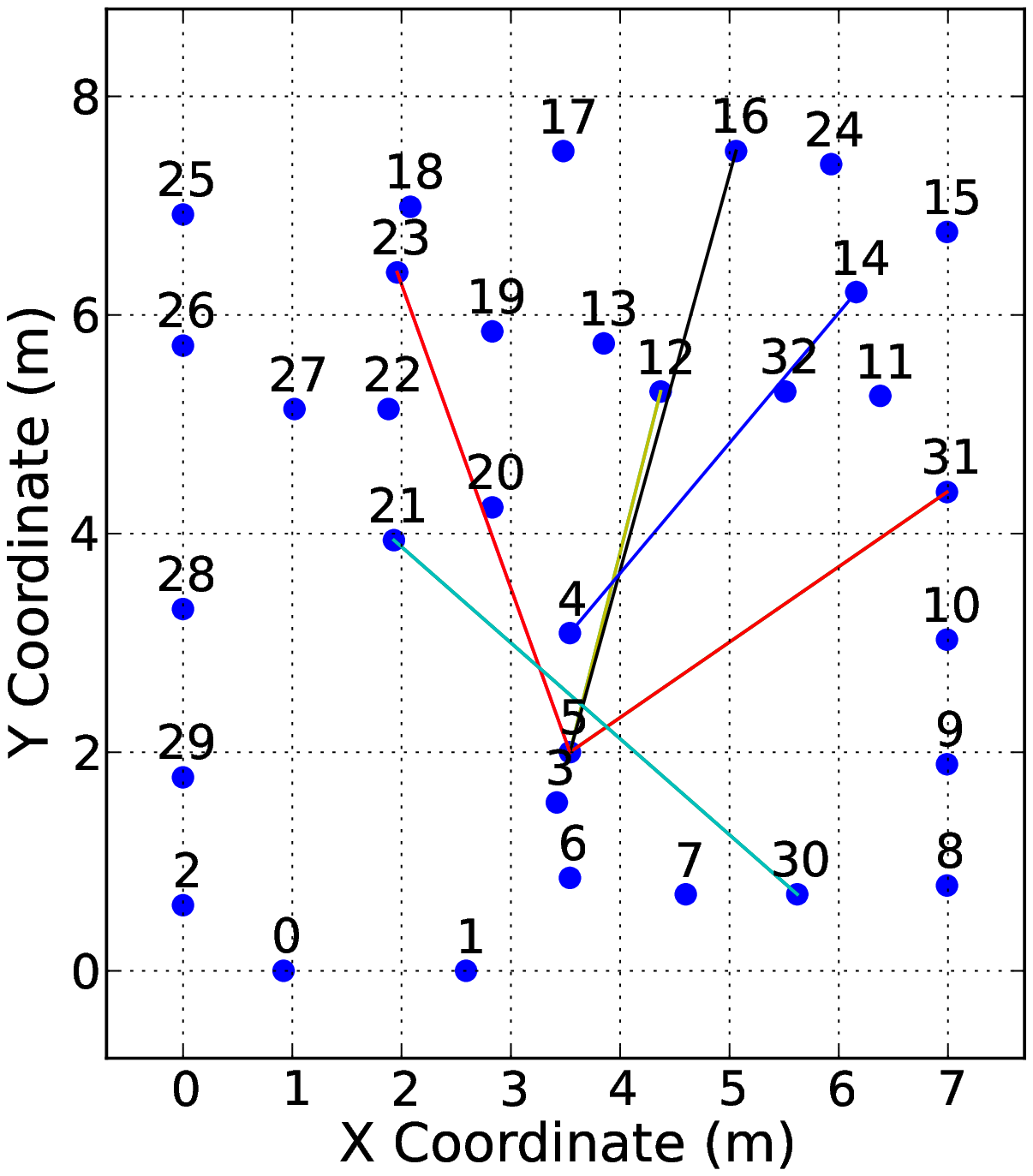,width=3.15in} 
             (b) \epsfig{figure=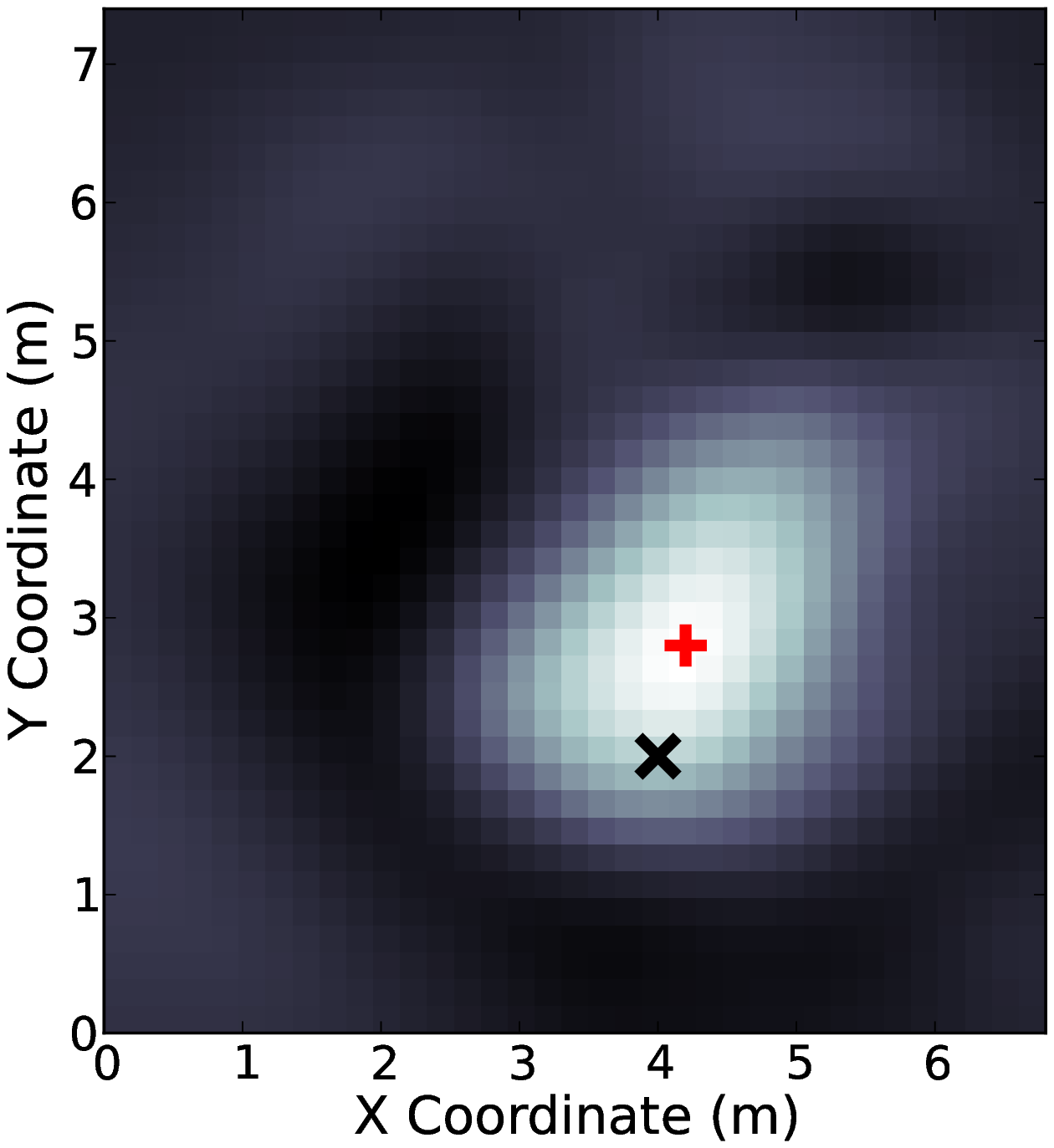,width=3.15in}}
\caption{(a) lines for the top ten links by $v_l$, and (d) breathing image $\hat{x}$ (white=high), with true ($\boldsymbol{\times}$) and estimated (\textcolor{red}{$\boldsymbol{+}$}) locations.}
    \label{F:breathingLocMaps}
\end{figure*}

We implement the breathing mapping and localization of (\ref{E:inversion}) using the parameters given in Table \ref{T:LocParameters}, and evaluate its performance using the apartment experiment.  Using the same example window as that shown in Figure \ref{F:RSSSignal_2}, we first draw link lines corresponding to the few links $l$ with highest $v_l$ in Figure \ref{F:breathingLocMaps}(b).  These include the four plotted in Figure \ref{F:RSSSignal_2}. 
The image $\hat{x}$ for the same data is shown in Figure \ref{F:breathingLocMaps}(c).  For this example, the location estimate has 0.82 m error.

\begin{table}[tbp]
\centering
\begin{tabular}{|l|cc|}
\hline
\bf          & \multicolumn{2}{c|}{\bf Avg.~Loc.~Error (m)} \\
\bf Test Loc &  \bf Basic & \bf Breakpoint \\
\hline
 Sofa       & 1.5 & 1.7 \\ 
 Table      & 2.1 & 2.1 \\ 
 Kitchen    & 2.9 & 3.6 \\
 Bathroom   & 1.5 & 1.9 \\
 Bed        & 2.6 & 2.7 \\
\hline
\bf Average & 2.1 & 2.4\\ 
\hline
\end{tabular}
\caption{Average location errors for basic and breakpoint methods} \label{T:LocResults}
\end{table}

We summarize the performance of the localization estimator across all windows and tests in the apartment experiment in Table \ref{T:LocResults}.  Using the basic method for mean removal, the average localization error over all tests is 2.1 m.  In a 7 by 8 meter area, this is a somewhat coarse estimate, however, it is typically sufficient to tell which room the person is in. Counterintuitively, for the breakpoint method, the localization estimates degrade somewhat to 2.4 m.  From examination of the RSS data and images produced using the basic method for mean removal, we can see that when windows include motion interference caused by the experimenter, the breathing map $\hat{x}$ can be as good or better than without the motion interference.  This is because the experimenter is causing the motion, which then leaks into the RSS signal on links near the experimenter, and helps to increase the image value near the experimenter's location.  The breakpoint method removes much of this motion interference, which then 
prevents it from increasing the 
image value near the correct location.

We notice some times when the image suddenly shifts and shows a bright spot at the top of the apartment, where there is a hallway outside of the apartment wall.  The experimenter had heard people walking through the hallway during some tests.  However, we did not record the ground-truth times of these events and thus future controlled experiments are suggested.

Finally, we note computation time is about 0.2 s per 30 s window, for both rate estimation and localization, in Python on a two-core 2.0 GHz processor -- thus real time monitoring is very possible.

\subsection{Nap Experiment Results}

\begin{figure}[htbp]
\centerline{ \epsfig{figure=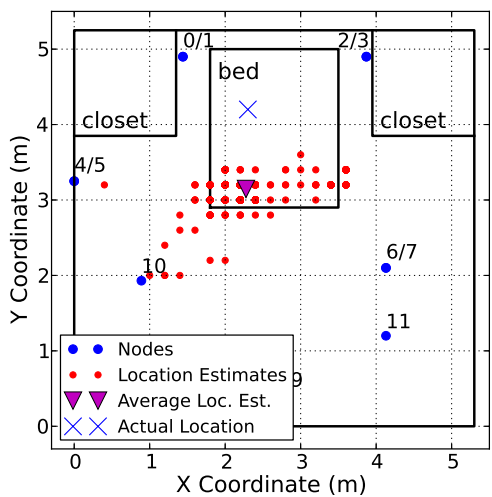,width=3.8in}}
\caption{Location estimates for each 30 second window (of which there are 792) during the nap experiment, along with an average over all windows, and the actual coordinate of the center of the person's chest during the nap.  }
    \label{F:breathingLocMapsNap}
\end{figure}

Finally, we test the nap experiment data for localization performance.  We use the same algorithm and parameters.  There is only one actual coordinate, which we measure to be the position of the chest when the person started napping.  There are 792 estimates, which are shown in Figure \ref{F:breathingLocMapsNap}.  Since we use the center of the maximum pixel of $\hat{x}$ as the location estimate, many of the location estimates exactly overlap.  We also show the average of all location estimates in Figure \ref{F:breathingLocMapsNap}.  This average is about 1.0 m away from the true chest position.  Considering the estimates individually, the RMS localization error is 1.16 m.   In general, performance is better than the average of the apartment experiment, but the size of the area is also smaller as well, about 1/3 of the area of the apartment.

\section{Performance vs.~Size}

In this section, we explore the performance of breathing monitoring when a smaller number of sensors, or a smaller number of channels, or both, are used in the network.  From a system design perspective, it would be beneficial to deploy a smaller number of nodes, and for those nodes to make as few measurements as possible.  If fewer channels were required to be measured, then one might transmit and receive less, thus enabling lower energy consumption, and less interference to other devices.

\subsection{Number of Nodes}

In this section, we show the performance of breathing monitoring when we only use a subset of nodes, to test using experimental data how the performance changes as a function of $S$.  For this, we use the data from the nap experiment. We would like to know, how many nodes are required for acceptable performance?  

Since we do not have a ``ground truth'' breathing rate for the nap experiment, we evaluate the performance of breathing rate estimation as follows.  We assume that short-term variation in breathing rate estimates are characteristic of a less-accurate estimator.  As such, we report the RMS difference between a current breathing rate and the short-term median, where the median is taken from all breathing rates estimated in a 90-second window centered at the current time, which we refer to as the ``RMS-median''.  

\begin{figure}[htbp]
\centerline{ \epsfig{figure=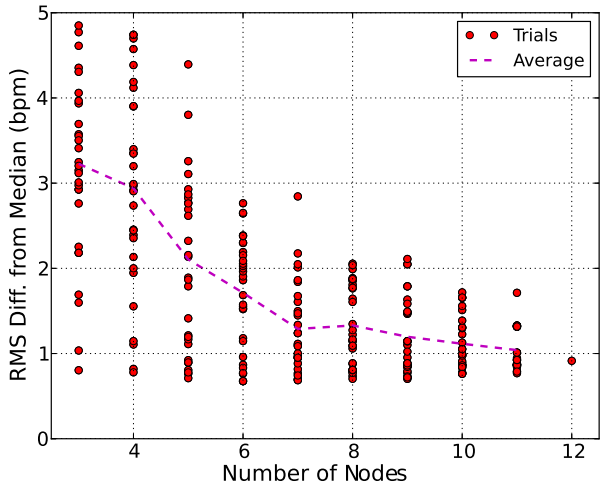,width=3.7in}}
\caption{RMS of the difference between the estimate and the short-term median, as a function of $S$, the number of nodes, for nap experiment.  }
    \label{F:rms_diff_from_median_random_subsets}
\end{figure}

First, we test using random subsets of the twelve nodes.  For each number of included nodes $S<12$, we run 30 random trials.  We show the RMS-median value for each trial, as well as the average vs. $S$, in Figure \ref{F:rms_diff_from_median_random_subsets}.  The RMS-median value, on average, decreases with $S$; however, the lowest possible RMS-median value seems to have a shallow minimum at a value of $S=6$, and every tested subset size had some trial with RMS-median value below 1.0.  It would seem that low number of nodes is fine if you select the \emph{right} nodes.

To explore this further, we test specific (non-random) subsets of nodes and compare the results in Table \ref{T:SubsetResults}.  First, we have eight nodes near the four corner of the bed, with nodes 0, 2, 4, and 6 at a height of about 0.67 m, and nodes 1, 3, 5, 7 on the floor.  We first test the four nodes on the floor, $\{1,3,5,7\}$, and see an RMS-median of 1.96 bpm.  In contrast, the four nodes at 0.67 m, $\{0,2,4,6\}$ have RMS-median of 0.72 bpm.  We suspect that the links between nodes on the floor have signals which propagate largely along the floor, and as such do not interact strongly with the person in the bed.  We also test using nodes $\{0,1,2,3\}$, the nodes closest to the person.  Using this set of nodes results in a very high RMS-median of 3.57 bpm.  Although they are close to the person, they are close to the person's head, and no link line crosses the person's chest.  

\begin{table}[tbp]
\centering
\begin{tabular}{|cc|}
\hline
\bf Subset &  \bf RMS-median \\
\bf of Nodes & \bf (bpm) \\
\hline
$\{0,2,4,6\}$       & 0.72 \\ 
$\{1,3,5,7\}$       & 1.96 \\ 
$\{0,2,4,6,8\}$       & 0.84 \\ 
$\{0,1,2,3\}$       & 3.57 \\ 
$\{0,1, \ldots, 5\}$       & 1.31 \\ 
$\{0,1, \ldots, 7\}$       & 0.68 \\ 
$\{0,1, \ldots, 9\}$       & 0.73 \\ 
$\{0,1, \ldots, 11\}$       & 0.91 \\ 
\hline
\end{tabular}
\caption{RMS-median using node subsets for nodes numbered as in Figure \ref{F:map_nap_experiment}. } \label{T:SubsetResults}
\end{table}

\subsection{Number of Channels} \label{S:NumberOfChannels}

\begin{figure}[htbp]
\centerline{ \epsfig{figure=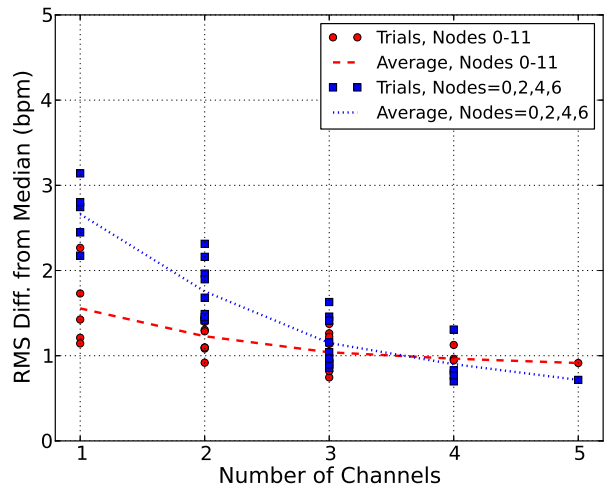,width=3.7in}}
\caption{RMS of the difference between the estimate and the short-term median, as a function of $C$, the number of channels, for the nap experiment.  The results for using all twelve sensors $\{0, \ldots, 11\}$, as well as using the node subset $\{0,2,4,6\}$, are both shown. }
    \label{F:nap_rmsdm_subset_channels}
\end{figure}

In prior work \cite{patwari11breathing}, nodes used only one channel.  We are interested in whether our performance has improved due to the measurement of multiple channels.  Here, we test the performance during the nap experiment when only links from a subset of the channels are used.  We test each possible (non-empty) subset of the five measured channels, and plot the RMS-median performance in Figure \ref{F:nap_rmsdm_subset_channels}.  First, using all $S=12$ nodes, we see average performance improves as the number of utilized channels increases, although slowly.  At $C=1$, the average RMS-median is 1.56 bpm, while when using all $C=5$ channels, the value is 0.91, a 41\% reduction.  The result is much more pronounced when re-running the test using only the nodes $\{0,2,4,6\}$.  These were a set of four sensors at height 67 cm closest to the bed, which achieve a RMS-median of 0.83 when using all $C=5$ channels.  The improvement for this set, when increasing from one to four channels, is dramatic.  The 
average RMS-median with $C=1$ for this set of nodes is 2.54 -- thus five channels achieves a 73\% reduction.  

Interestingly, in some cases, a lower number of channels and a lower number of nodes can actually improve performance.  We do not believe that it is better, on average, to use fewer sensors or fewer channels, because we may not know ahead of time which sensor positions will be best, and we can't know ahead of time which set of channels is best.  Regardless, future work may exploit adaptive methods to select channels from among those that can be measured, or adapt which set of sensors should be operating, in order to best estimate breathing rate.  Such adaptive protocols could dramatically reduce the energy used in a breathing monitoring RF sensor network.  Further, it would seem that better algorithms, which are more aligned with the statistics of the measured data, should be developed so that it doesn't \emph{degrade} performance when more links' data is included.

\section{Conclusion} \label{S:Conclusion}

This paper demonstrates key advances for the use of RSS in a wireless network to monitor the breathing of a person in the deployment area.  First, we introduce a new method for mean removal that makes the method robust to small movements, a key requirement to be able to perform RSS-based breathing monitoring in real-world residential scenarios.  Second, we show that the link PSD at the breathing rate can be used to estimate the location of the breathing person.  We test both advances using a multi-channel network deployed in an apartment building.  The results show the possibilities for accurate and reliable breathing monitoring and localization of a breathing person for a variety of applications.  

The area of wireless network breathing monitoring has many unanswered questions.  What are good statistical and radio propagation models that explain a link's ability to measure breathing as a function of the person's position?  Can adaptive algorithms and protocols be used to reduce the sampling requirements, and thus reduce the traffic and energy required for breathing monitoring?  How should breathing be measured with other wireless hardware, such as 802.11 devices?  What algorithms should be used to monitor and track the breathing of multiple people in the same deployment area?  What WLAN protocols could be used to prevent an adversary from surreptitiously using someone's wireless network to eavesdrop on their breathing?  Each of these questions may result in interesting and useful research directions.

\section*{Acknowledgements}
The authors would like to thank Brad Mager, who helped with the experiments.



\end{document}